\documentclass[12pt]{article}

\def\beq{\begin{eqnarray}}
\def\eeq{\end{eqnarray}}
\def\bea{\begin{eqnarray}}
\def\eea{\end{eqnarray}}

\newcommand{\gsim}{\lower.7ex\hbox{$\;\stackrel{\textstyle>}{\sim}\;$}}
\newcommand{\lsim}{\lower.7ex\hbox{$\;\stackrel{\textstyle<}{\sim}\;$}}

\newcommand{\drawsquare}[2]{\hbox{%
\rule{#2pt}{#1pt}\hskip-#2pt%  left vertical
\rule{#1pt}{#2pt}\hskip-#1pt%  lower horizontal
\rule[#1pt]{#1pt}{#2pt}}\rule[#1pt]{#2pt}{#2pt}\hskip-#2pt%  upper horizontal
\rule{#2pt}{#1pt}}% right vertical

\newcommand{\fund}{\raisebox{-.5pt}{\drawsquare{6.5}{0.4}}}%  fund
\newcommand{\Ysymm}{\raisebox{-.5pt}{\drawsquare{6.5}{0.4}}\hskip-0.4pt%
        \raisebox{-.5pt}{\drawsquare{6.5}{0.4}}}%  symmetric second rank
\newcommand{\Yasymm}{\raisebox{-3.5pt}{\drawsquare{6.5}{0.4}}\hskip-6.9pt%
        \raisebox{3pt}{\drawsquare{6.5}{0.4}}}%  antisymmetric second rank
\newcommand{\antifund}{\overline{\fund}}

\begin{document}

\setlength{\baselineskip}{0.25in}

%#!latex

%\twocolumn[\hsize\textwidth\columnwidth\hsize\csname
%@twocolumnfalse\endcsname
%%
%%
\begin{titlepage}
\noindent
\begin{flushright}
MCTP-05-83 \\
\end{flushright}
\vspace{1cm}

\begin{center}
  \begin{Large}
    \begin{bf}
Surveying Standard Model Flux Vacua on $T^6/Z_2\times Z_2$

    \end{bf}
  \end{Large}
\end{center}
\vspace{0.2cm}
\begin{center}
\begin{large}
Jason Kumar and James D. Wells \\
\end{large}
  \vspace{0.3cm}
  \begin{it}
Michigan Center for Theoretical Physics (MCTP) \\
        ~~University of Michigan, Ann Arbor, MI 48109-1120, USA \\
\vspace{0.1cm}
\end{it}

\end{center}

%\maketitle
\begin{abstract}

We consider the  $SU(2)_L\times SU(2)_R$
Standard Model brane embedding in an orientifold of $T^6/Z_2
\times Z_2$.  Within defined limits, we construct all
such Standard Model brane embeddings and determine the relative
number of flux vacua for each construction.  Supersymmetry preserving
brane recombination in the hidden sector enables us to identify many
solutions with high flux. We
discuss in detail the phenomenology of one model which is likely
to dominate the counting of vacua.  While K\"ahler moduli
stabilization remains to be fully understood, we define the criteria
necessary for generic constructions to have fixed moduli.

\end{abstract}

\vspace{1cm}

\begin{flushleft}
hep-th/0506252 \\
June 2005
\end{flushleft}

\end{titlepage}
%\pacs{PACS numbers: }
%]

\setcounter{footnote}{1}
\setcounter{page}{2}
\setcounter{figure}{0}
\setcounter{table}{0}

\tableofcontents

\section{Introduction}

There is now substantial, though by no means conclusive,
evidence that there are a very large number of
vacua of string theory in which all\cite{Kachru:2003aw}
of the moduli\cite{flux1,flux2,flux3,Bousso:2000xa,flux4,flux5} can be fixed.
Recently, there has been significant development in the
techniques for describing and counting these vacua
\cite{Douglas:2003um,Ashok:2003gk,Denef:2004ze,describeandcount,IIA}.
Much of this work is statistical in
nature, generating estimates of the number of vacua with
specific gross properties
\cite{Bousso:2000xa,grossprop}.  The actual construction of explicit
models with all moduli vevs and their potentials identified
is far more difficult\cite{Denef:2004dm,Robbins:2004hx,
Denef:2005mm}, though constructions using Type IIA string
compactifications seem more tractable than those of
Type IIB\cite{IIA}.

Our interest in this work is to gain insight into the statistical distributions
of vacua and their properties, and in particular how that can be connected to the
problem of model-building\cite{Marchesano:2004yq,Marchesano:2004xz,Dijkstra:2004cc,
Blumenhagen:2004xx,Cvetic:2005bn}.  It seems clear that any concrete
program for actually constructing stringy models of the
real world (or even of toy models which resemble the world and
have interesting phenomenology) can
only be helped by a statistical study of
the frequency with which low-energy properties occur on
the landscape.

One direction is to characterize the distribution of vacua
compatible with the standard model (SM)
\cite{Marchesano:2004yq,Marchesano:2004xz,IDB,Cvetic:2005bn}.
This would be a difficult
endeavor, even for the limited case of orientifolded Calabi-Yau
three-fold compacitifications of Type IIB string theory.  We
will instead consider as an exercise the simple construction
of Type IIB compactified on an orientifold of
$T^6 /Z_2 \times Z_2$.  We hope the results of this survey
can provide some intuition about the more general problems of constructing
SM string embeddings on the landscape, and of determining
the distribution of these embeddings.

In these constructions, some gauge dynamics will arise from open
strings beginning and ending on branes.
We would like this open string gauge theory to include the
SM.  As a result, we can roughly divide the open string
gauge theory into two sectors: the visible sector containing the
SM gauge group (with some extensions, such as to a
Pati-Salam unification group or a left-right extension), and a
hidden sector containing gauge groups not identified with the SM.
The branes that are relevant for visible or hidden sector dynamics
will wrap some holomorphic even-dimensional cycles of the Calabi-Yau,
and will be extended in all of the non-compact directions.

One must first define what is meant by a string
construction of the SM.  We will consider a series
of branes embedded in an orientifolded Calabi-Yau 3-fold
compactification such that one set of branes yields the gauge
group and chiral matter content of the SM.  This set
of branes will be called the visible sector.  Various additional
hidden sector branes will also be allowed, and there may be chiral
exotics charged under either hidden sector branes, or both hidden
and visible sector branes.  The only demand we will make of the
matter content is that there be no chiral exotics charged only under
the visible sector.  In particular, note that there is no restriction
of any kind on vector-like matter, as this matter can receive a large
mass and thus not conflict with experiment.

In section 2, we will review the general properties of our $T^6/Z_2
\times Z_2$ orientifold.  In section 3, we will codify the rules for
constructing consistent brane embeddings.  In section 4, we discuss
actual SM embeddings, and the properties of associated
flux vacua.  In section 5, we discuss a model which dominates the
counting of SM flux vacua.  We close with a discussion
of our results in section 6.

%%%%%%%%%%%%%%%%%%%%%%%%%%%%%%%%%%%%%%%%%%%%%%%%%%%%%%%%%%%%%%%%%%%%
\section{$T^6/Z_2\times Z_2$ Orientifold}

We will focus on brane constructions of the SM on
an orientifold of $T^6 /Z_2 \times Z_2$.
\cite{Marchesano:2004yq,Marchesano:2004xz,Gopakumar:1996mu,T6Z2Z2}.
The $Z_2 \times Z_2$ orbifold group is thus
supplemented by the orientifold element $\Omega R$;
the orbifold group is generated by the elements

\bea
\alpha &:& (z_1 ,z_2 ,z_3) \rightarrow (-z_1 ,-z_2 ,z_3)
\nonumber\\
\beta &:& (z_1 ,z_2 ,z_3) \rightarrow (z_1 ,-z_2 ,-z_3)
\eea
where the involution is given by
\bea
R &:& (z_1 ,z_2 ,z_3) \rightarrow (-z_1 ,-z_2 ,-z_3)
\eea
and $\Omega$ is the worldsheet parity operator.

This orientifold group generates 64 O3-planes and 12 O7-planes.
The O3-planes stretch in the 4 non-compact directions.  The
O7-planes appear in three sets which all stretch in the non-compact
directions, but which wrap different 4-cycles in the compact directions.
In particular, the O7-planes respectively wrap the tori $T_1 ^2 T_2 ^2$,
$T_1 ^2 T_3 ^2$ and $T_2 ^2 T_3 ^2$.  The $T^6/Z_2 \times Z_2$ orbifold
itself breaks $N=8$ supersymmetry down to $N=2$.  The orientifold action
further breaks supersymmetry down to $N=1$.

For our choice of discrete torsion, this model has
3 K\"ahler moduli (which determine the size of each $T^2$ factor)
and 51 complex sturcture moduli.  48 of these
complex structure moduli arise
at the fixed points of elements of the orbifold group.
For a different choice of
discrete torsion, the 48 moduli arising at the fixed points would
instead be K\"ahler moduli.

In this compactification, branes may wrap either the cycles
of the $T^6$, or the shrunken cycles arising at the orbifold fixed
points.  We will only be interested in the cycles of the $T^6$.  We
may describe the relevant branes of Type IIB string theory (D3-,D5-,D7- and
D9-branes) in the so-called magnetized D-brane
formalism\cite{Cascales:2003zp}.  This formalism
arises from the realization that lower dimensional branes can be
described by magnetic fluxes on the worldvolume of a D9-brane.  Essentially,
we factorize the six-torus into $(T^2)^3$, and we assign to each brane
an ordered pair $(n_i ,m_i)$ for each $T_i ^2$, where $m_i$ is the number
of times that the brane wraps this $T^2$, and $n_i$ gives the amount of
constant magnetic flux on this cycle normalized as

\beq
{m_i \over 2\pi} \int_{T_i ^2} F^i =n_i
\eeq

We see that a brane in which 3 of the $m$'s are zero must be a D3-brane
extended in the non-compact dimensions, but with no wrapping on the torus.
Similarly, a D5-brane wrapping a 2-cycle of the compact space will have only
one non-zero $m$, while a D7-brane wrapping a 4-cycle of the compact space
will have two non-zero $m$'s.  For a D9-brane, all $m$'s will be non-zero.

It is often easier to picture this from the T-dual prescription,
which is Type IIA string theory on a $T^6/Z_2 \times Z_2$ orientifold.  In
this picture, the involution $R'$ which appears in the orientifold element
$\Omega R'$ is given by

\beq
R': (z_1, z_2, z_3) \rightarrow (z_1 ^*, z_2 ^*, z_3 ^*)
\eeq
The IIB branes are dualized to D6-branes of
Type IIA which wrap a one-cycle of each torus.  The $(n_i, m_i)$
ordered pairs just give the winding numbers on each cycle of
the torus $T^2 _i$.

If $\Delta_a$ is any particular brane, its
orientifold image $\Delta'_a$ will
have wrapping numbers given by $n^a_i \to n^a_i$,
$m^a_i\to -m^a_i$.  It is easiest to see this by examining the
T-dual IIA picture, in which the $m$'s are the wrapping numbers
along the three real directions which are inverted by the involution
$R$.

%%%%%%%%%%%%%%%%%%%%%%%%%%%%%%%%%%%%%%%%%%%%%%%%%%%%%%%%%%%%%%%%%%%%%%
\section{Rules for Brane Constructions}

The rules for constructing brane models in type IIB theories
are detailed in the literature.  We wish here to distill these
discussions into a list of rules that must be followed to
construct D-brane models in our setup.  When all rules are
satisfied, a mathematically consistent theory results.

{\bf Wrapping numbers:}
In the previous section we discussed the salient properties
of the $T^6/Z_2\times Z_2$ orientifold, and showed that
a brane $\Delta_a$ could be written as a set of its wrapping numbers
on the three two-tori cycles
\beq
\Delta_a=(n^a_1,m^a_1)(n^a_2,m^a_2)(n^a_3,m^a_3).
\eeq
The $n^i$ and $m^i$ numbers must be co-prime integers.

{\bf RR tadpoles:}
Gauss's law imposes the constraint that there may be no net charge in a compact
space.  As a result, RR charges which stretch over
all non-compact directions
must cancel.  This constraint can be rephrased as the statement that,
for a consistent brane embedding, all RR tadpoles must cancel.
It is easy to implement these tadpole conditions by constructing
a brane-charge vector $\vec Q$ such that $Q_0$ is the $D3$ brane
charge, and $Q_i$ are the $D7_i$ brane charges\footnote{The minus signs in the
definitions of $Q_{1,2,3}$ serve the purpose of assigning $+1$ charge to
pure $D7_i$ branes, just as there is a charge of $+1$ to a pure $D3$ brane.
Note that $D7_i$ refers to a D7-brane which is not wrapped on the torus
$T_i ^2$, but is wrapped on the other two.}:
\bea
Q_0 & = & n_1n_2n_3 ~~ (D3~~{\rm charge})\nonumber\\
Q_1 & = & -n_1m_2m_3 ~~ (D7_1~~{\rm charge})\nonumber\\
Q_2 & = & -m_1n_2m_3 ~~ (D7_2~~{\rm charge})\nonumber\\
Q_3 & = & -m_1m_2n_3 ~~ (D7_3~~{\rm charge})
\eea

This charge vector, summing over
all branes $\Delta_a$ and their images $\Delta'_a$ (where $m^a_i\to -m^a_i$),
must equal the sum of all orientifold plane charges:
\beq
\sum_a N_a(\vec Q(\Delta_a)+\vec Q(\Delta'_a))=32 \vec Q({\cal O})
\eeq
where $N_a$ is the number of branes for each stack $a$.
Since $\vec Q({\cal O})=(1,1,1,1)$ and $\vec Q(\Delta_a)=\vec Q(\Delta'_a)$,
we have as the final condition
\beq
\sum_a N_a \vec Q(\Delta_a) = (16,16,16,16).
\eeq
Three-form fluxes will
contribute to the $D3$ charge but not the $D7_i$ charge. One unit
of flux contributes 32 units of $D3$ charge, thereby changing the RR
tadpole conditions to
\beq
\sum_a N_a \vec Q(\Delta_a) + (32N_{\rm flux},0,0,0) = (16,16,16,16).
\label{RR constraint}
\eeq
where $N_{\rm flux}$ is a non-negative integer, and again, this final sum is
only over the branes and not their images.  The factor of 32 arises from the
reduction in size of the 3-cycle volume due to the orbifold action
\cite{Blumenhagen:2003vr} and from the condition that there by
no exotic branes\cite{Frey:2002hf}.

It is interesting to note that the cancellation of RR tadpoles implies
the cancellation of anomalies in the worldvolume gauge theories of
the embedded branes\cite{Berkooz:1996km,Uranga:2000xp}.
Indeed, the only consistency conditions, from the string theory point of
view, will amount to tadpole cancellation conditions.  Consistency of the
overall string compactification implies consistency of the low-energy
description, although this may appear through an anomaly inflow which
cancels local anomalies.  This fact will have
important implications later for the number of chiral exotics in the embeddings
we consider.

{\bf K-theory constraints:}
The RR tadpole conditions are
requirements on the total $D3$ and $D7$ charges
of the brane stacks.  There are analagous constraints on the
total $D5$ brane and $D9$ brane charges
\cite{Cascales:2003zp}. We can form a new
vector $\vec Y$ of these charges, such that
\bea
Y_0 & = & m_1m_2m_3~~(D9~{\rm charge}) \nonumber\\
Y_1 & = & m_1n_2n_3~~(D5_1~{\rm charge}) \nonumber\\
Y_2 & = & n_1m_2n_3~~(D5_2~{\rm charge}) \nonumber\\
Y_3 & = & n_1n_2m_3~~(D5_3~{\rm charge})
\eea
The requirement on these charges is merely that they be
even under the summation
\beq
\sum_a R_a \vec Y^a = (0,0,0,0) ~{\rm mod}~2
\eeq
where $R_a$ is the rank of the gauge group for brane stack $a$.
In fact, this is equivalent to the demand that if
one inserts a probe D-brane with $SU(2)$ gauge group, then there
should be no global anomaly (i.e., the number
of Weyl fermions in the probe brane worldvolume theory must be even
\cite{Uranga:2000xp,Witten:1982fp}).
For brane stacks at orbifold fixed points, $R_a=N_a/2$ and the
constraint becomes
\beq
\sum_a N_a \vec Y^a = (0,0,0,0) ~{\rm mod}~4.
\label{K-theory constraint}
\eeq
This is often called the K-theory constraint, since it is a restriction
on the theory not captured by pure homology
\cite{Witten:1998cd}. It might appear rather
innocuous, and in many cases in the early literature it had no strong
impact on the results.  But in making theories with magnetized $D9$ branes,
this constraint can be quite restrictive, as we will encounter
in the next section.

It turns out that adding discrete $B$-field(s) along the tori
will enable us to more
easily find solutions to these K-theory constraints.  However,
these $B$-fields will also change the RR-tadpole constraints,
making them more difficult to solve.  In any case, they can potentially
introduce obstructions to the vector bundles which would be
necessary for us to obtain the open string gauge groups which
we will need for our SM construction\footnote{We
are indebted to S. Sethi and G. Shiu for discussions of these points.}.
As a result, we will not consider turning on any of these
discrete $B$-fields.

{\bf NSNS tadpoles:}
NSNS tadpole cancellation (i.e., no uncompensated brane tensions)
is achieved if the $N=1$ supersymmetry that remains from the orientifold
is preserved in all D-brane sectors.  Operationally, this leads to
the turning on of a Fayet-Iliopoulos (FI) term if the tadpole is not cancelled.
Each brane has a FI-term which is determined in terms of the set of
three K\"ahler moduli parameters
${\cal A}_1$, ${\cal A}_2$ and ${\cal A}_3$ by
the equation\footnote{This derives from the fact that two
branes preserve common supersymmetries if they can be related by
an $SU(d)$ rotation in $d$ complex dimensions\cite{Berkooz:1996km}.}
\beq
\sum_{i=1}^{3} \tan^{-1}(m^a_i{\cal A}_i,n^a_i)~{\rm mod}~2\pi=
\theta \sim \xi_{FI}
\label{supersymmetry condition}
\eeq
where $\alpha =\tan^{-1}(y,x)$ is defined from 0 to $2\pi$ (or
equivalently, $-\pi$ to $\pi$) in such a way that
$\sin \alpha = {y\over \sqrt{x^2 +y^2}}$ and
$\cos \alpha = {x\over \sqrt{x^2 +y^2}}$, and where ``$\rm mod$"
is defined such that $\theta \in [-\pi ,\pi]$.

It is important to note that if NSNS tadpoles are not 
cancelled~\cite{Kachru:1999vj},
supersymmetry will {\bf not} necessarily be broken.  Indeed, the
FI-term will appear in the $D$-term potential in the form

\beq
\label{Dtermpotential}
V_{D_a} = {1\over 2g^2} (\sum q_i |\phi_i |^2 +\xi)^2
\eeq
where the $\phi$ are scalars charged under the diagonal $U(1)$
of the gauge group associated with the stack of branes $\Delta_a$.
This implies that even if $\xi \neq 0$, an appropriate scalar can
possibly get a vev to cancel the $D$-term
potential.  This corresponds to brane recombination, in which the two
branes that bind are those under which the veved scalar is charged.
Thus, we should think of this FI-term as simply contributing to a real
constraint equation involving both open string moduli and the
K\"ahler moduli.  We will indeed find that our brane embeddings will
necessarily be very rich, yielding many scalars charged under each
$U(1)$ with both positive and negative sign.  As such, it is most
likely that
a non-vanishing FI-term will in fact lead to a deformation of the
brane system which preserves supersymmetry.

However, we should distinguish between branes that satisfy the NSNS tadpole
conditions somewhere in K\"ahler moduli space, and those that cannot
satisfy them anywhere in the moduli space (such as $\overline{D3}$ branes, for example).
Branes that cannot satisfy the tadpole conditions anywhere in moduli space
have the potential to destabilize the solution.  An example of such an
instability was shown in \cite{Kachru:2002gs}, where it
was found that the presence of sufficiently many anti-D3 branes would result
in a classical instability through which $\overline{D3}$-branes and fluxes would
annihilate.  More generally, $\overline{D3}$-branes will also contribute a term to 
the potential
arising from their vacuum energy, which can destabilize the K\"ahler
moduli of the solution unless
it is tuned to be small.  What distinguishes an $\overline{D3}$-brane from a supersymmetric
brane (such as a D3-brane) is its orientation.  In our previous language, the
$\overline{D3}$-brane maximally violates the NSNS tadpole constraint, and cannot
be made to satisfy it anywhere in K\"ahler moduli space.

On the other hand, 
a brane that can satisfy the NSNS tadpole conditions somewhere will indeed
be supersymmetric and stable (with unbroken gauge symmetry) when the FI-term
vanishes.  Therefore, the only potentially destabilizing contribution which they
make to the potential is from the FI-term, and in fact this term merely provides
another constraint which can be satisfied (restoring supersymmetry).

Although we do not have a general argument that branes that cannot satisfy the
NSNS tadpole conditions anywhere in K\"ahler moduli space will lead to instability,
our experience with $\overline{D3}$-branes leads us to strongly suspect
that such branes have the potential to destabilize the solution.
As such, we will demand that all branes be able to satisfy the NSNS tadpole
constraints somewhere in K\"ahler moduli space, but not necessarily at the same point.
This will ensure that our solutions are likely to be truly stable, though our limitation
may exclude some stable solutions.
Indeed, one may nevertheless be able to add a small number of $\overline{D3}$ branes to such
solutions for the purpose of breaking supersymmetry.  However, one must then
be careful to tune their contribution to avoid destabilizing the solution.
We will not make this choice rigid, in order to preserve the
potentially more attractive possibility of a
different mechanism for supersymmetry breaking (such as IASD fluxes).  We
will discuss the issue of supersymmetry breaking in more detail in
sec.~\ref{phenomenological considerations}.

{\bf Gauge groups of brane stacks:}
We will choose brane stacks to either lie on orientifold planes or on
orbifold fixed points away from orientifold planes. This will ensure
that we have an odd number of generations~\cite{IDB}.
The resulting gauge groups are different in these two cases.

For a stack with $N_a$ branes on
an orientifold plane\footnote{Sometimes, $N_a$ branes on an orientifold brane
are identified with their images, leading some authors to say by convention that
there are $2N_a$ branes
for this case.}, the gauge group
\footnote{There is some freedom in the choice of action of the orientifold
group on the Chan-Paton indices.  Indeed, Denef et al.\cite{Denef:2005mm} 
use this freedom to construct (non-standard
model) embeddings with $SO(N)$ gauge groups.  One can consider whether
these discrete choices allow one to find SM constructions,
and how these constructions relate to the ones we consider.}
is $USp(N_a)$.  This gauge group
only makes sense if $N_a$ is even.  ($USp(2)$ is isomorphic to $SU(2)$,
which we will use below.)
Only pure $D7$ branes and $D3$ branes
can be located on orientifold planes and give rise to these $USp(N_a)$
gauge groups. For a stack with $N_a$ branes at an orbifold fixed point, but not
on an orientifold plane, the gauge group rank is $U(N_a/2)$. Again, $N_a$
must be even.

{\bf Intersection numbers and chiral matter content:}
The chiral matter content is obtained by computing intersection numbers
of one brane stack with another:
\beq
I_{ab}=\prod_{i=1}^3 \det \left(
\begin{array}{cc} n^a_i & m^a_i \\ n^b_i & m^b_i \end{array}\right)
=(n^a_1 m^b_1 - m^a_1 n^b_1)(n^a_2 m^b_2 - m^a_2 n^b_2)
(n^a_3 m^b_3 - m^a_3 n^b_3)
\eeq

Let us label $a$ as a stack of a magnetized branes
(either magnetized $D7$ or $D9$).  The matter content arising from a
string beginning and ending on $a$ will be
\beq
aa~{\rm matter}: ~U(N_a/2)~{\rm vector~multiplet~plus~3}~{\rm adjoint~chirals}
\eeq
Furthermore, $a$ and its image $a'$ ($m^b_i\to -m^b_i$)
can intersect with any other brane $c$ and its orientifold
image $c'$ (applicable
only if $c$ is magnetized $D7$ or $D9$-brane stack):
\bea
ac~{\rm matter}: & ~ & I_{ac}~{\rm copies~of~}(\fund_a,\antifund_c)~{\rm chirals}
\nonumber\\
aa' ~{\rm matter}: & ~ & -(2I_{a,{\cal O}} -I_{aa'}/2)~{\rm copies~ of}~\Ysymm~{\rm chirals}
\nonumber\\
 & & -(2I_{a,{\cal O}} + I_{aa'}/2)~{\rm copies~ of}~\Yasymm~{\rm chirals}
\nonumber\\
ac' ~{\rm matter}: & ~ & I_{ac'}~{\rm copies~ of}~ (\fund_a,\fund_c)~{\rm chirals }
\eea
where $I_{a,{\cal O}}$ is the intersection number summed over the
each orientifold plane:
\beq
I_{a,{\cal O}}=m^a_1n^a_2n^a_3+n^a_1m^a_2n^a_3+n^a_1n^a_2m^a_3-m^a_1m^a_2m^a_3
\eeq

Let us label $b$ as a stack of pure $D7$ or $D3$ branes lying on an orientifold plane.
The matter arising from strings beginning and ending on $b$ are
\beq
bb~{\rm matter}: ~USp(N_b)~{\rm vector~multiplet~plus~3}~{\rm copies~of}~\Yasymm~{\rm chirals}
\eeq
$b$ can also intersect with any other brane
$c$ to yield
\beq
bc~{\rm matter}: I_{bc}~{\rm copies~of~}(\fund_b,\antifund_c)~{\rm chirals}
\eeq
There is no intersection of $b$ with its image or with the image
of any other brane that contributes more to the total matter content.

%%%%%%%%%%%%%%%%%%%%%%%%%%%%%%%%%%%%%%%%%%%%%%%%%%%%%%%%%%%%%%%%%
\section{Standard Model Embeddings}

One goal of a string model building exercise is to construct a
visible sector that has a chance of reducing to the SM
at low energies.
For us, this will mean a visible sector that contains the SM
gauge group with the correct chiral matter.
We need four different brane stacks in order to achieve a SM embedding:
a stack for each of the three gauge groups of the SM, plus another
stack to enable $SU(3)\times SU(2)_L$ singlets to intersect
with hypercharge.  Not only would we like these stacks to yield a
gauge group containing the SM with the right chiral
matter content, but we would also like this visible sector to
preserve the same $N=1$ supersymmetry as the orientifold.

Requiring the visible sector to contain the SM with 3
generations forces us to include visible sector branes
with large charges.  This will (in known constructions) mean that
the visible sector includes branes with $D3$-brane charge larger than
that carried by the $O3$-planes.  To compensate for this, one must
include a source of negative $D3$-brane charge.
One approach is to allow
anti-$D3$ branes into the spectrum, thereby breaking supersymmetry but
allowing large fluxes that stabilize the complex structure moduli.
A second approach is to construct a visible sector entirely out of
$D7$ branes and then using magnetized $D9$ branes with induced negative
$D3$ brane charge in the hidden sector to cancel the RR tadpole.  A
third approach is to make the magnetized $D9$ branes part of the visible
sector, and use only
pure $D7$ and $D3$ branes in the hidden sector to cancel the RR tadpoles.

All of these approaches to finding a SM embedding for type IIB are
laudable.  However, we wish to focus on the second approach
since it has the advantage of an economical gauge
group structure of the visible sector (i.e., $USp(2)$ groups rather than
$U(2)$ groups can generate the weak $SU(2)$) and allows a wide
variety of supersymmetry breaking mechanisms and scales without destabilizing
the solution.

The simplest structure for such a four-stack embedding is the left-right
model, as advocated by\cite{Marchesano:2004yq,Marchesano:2004xz}.
In this model, the $SU(2)_{L,R}$ groups arise from $USp(2)$ gauge
theories living on stacks of branes, rather than from $U(2)$.  This
feature can be quite attractive, as otherwise there will be additional
$U(1)_{L,R}$ anomalies which must be cancelled.  Of course it is
possible to cancel such anomalies, either in a strictly field theoretic
context through the use of $U(2)$ anti-doublets as part of the chiral
matter\cite{Ibanez:2001nd}, or in a string theory
context through the Green-Schwarz
mechanism.  However, the use of anti-doublets may not be desirable
from a phenomenological standpoint.  Although these difficulties can
be solved, they are avoided altogether in models which contain $USp(2)$ groups
in the visible sector.

We will consider only Pati-Salam left-right constructions
in which the $SU(2)_{L,R}$ groups arise from $USp(2)$ groups.  As a result,
two of the four brane stacks (those generating $U(3),U(1)\subset U(4)$)
have the same
wrapping numbers, while the other two stacks are either pure D3-branes or
pure D7-branes.  The
two $SU(2)$ branes must have different intersections with respect to the
$U(3)$ brane (to account for the chiral bifundamental matter).
As a result, we must
pick the $SU(2)$ stacks from two distinct choices out of the four pure
D3/D7-branes
\bea
D7_1 & (1,0)(0,1)(0,-1) \nonumber\\
D7_2 & (0,1)(1,0)(0,-1) \nonumber\\
D7_3 & (0,1)(0,-1)(1,0) \nonumber\\
D3 & (1,0)(1,0)(1,0)
\eea
There are thus six choices we can make, but without loss of generality,
we will choose $D7_2$ and $D7_3$\cite{Marchesano:2004yq,Marchesano:2004xz}.
Similar statements will apply for the other cases.

Since the $U(3)$ and $U(1)$
brane stacks have the same wrapping numbers, the only choice we
have left is with these six wrapping numbers.  These are chosen subject
to the constraint that the brane be supersymmetric somewhere in K\"ahler
moduli space, and that it have the right intersection numbers with the
two $SU(2)$ branes.  These intersection conditions can be rephrased as

\bea
n_1 m_2 n_3 &=& 3
\nonumber\\
n_1 n_2 m_3 &=& -3
\eea
If we wish to have no magnetized D9-branes in the visible sector, then
must have $n_1=1$,$m_1=0$.  We see then our
only choices
for the wrapping numbers of the $U(4)$ stack are
\bea
{\rm choice ~1}~b_{1,2,3}=0 & (1,0)(3,1)(3,-1) \nonumber\\
{\rm choice~ 2}~b_{1,2,3}=0 & (1,0)(1,3)(1,-3)
\eea

Thus, going with choice 1, we can identify the full visible sector of a three-generation
model of this type as four stacks of $D7$ branes
\bea
N_a=6 & (1,0)(3,1)(3,-1) \nonumber\\
N_b=2 & (0,1)(1,0)(0,-1) \nonumber\\
N_c=2 & (0,1)(0,-1)(1,0) \nonumber\\
N_d=2 & (1,0)(3,1)(3,-1)
\label{visible sector}
\eea
Gauge groups for this are $SU(3)\times SU(2)_1\times SU(2)_2\times U(1)_a\times U(1)_d$,
which yields the matter content given by table~\ref{SM Matter content}.  The $H_u$
and $H_d$ fields are vector complements that do not contribute to the
net chirality and are thus not accounted
for by the
intersection numbers between branes~\cite{Marchesano:2004xz}.  They are
charged states in the  $(bc)$ sector
despite $I_{bc}=0$, and we include them in the table
to provide a complete picture of what is the minimum visible sector allowed
by this framework consistent with SM needs.

\begin{table}[tbh]
\begin{center}
\begin{tabular}{cccccc}
\hline\hline
Sector & $N_{\rm copies}$ & $SU(3)\times SU(2)_L\times SU(2)_R$ & $U(1)_a$ &
$U(1)_d$ & $U(1)_{B-L}$ \\
\hline
$(ab)$ & 3 & $(d_L,u_L)\sim$ (3,2,1) & 1 & 0 & $1/3$ \\
$(ac)$ & 3 & $(d^c_R,u^c_R)\sim$ ($\bar{3}$,1,2) & $-1$ & 0 & $-1/3$ \\
$(db)$ & 3 & $(l_L,\nu_L)\sim$ (1,2,1) & 0 & 1 & $-1$ \\
$(dc)$ & 3 & $(l^c_R,\nu^c_R)\sim$ (1,1,2) & 0 & $-1$ & 1 \\
$(bc)$ & 1 & $(H_u,H_d)\sim$(1,2,2) & 0 & 0 & 0 \\
\hline\hline
\end{tabular}
\end{center}
\caption{Minimal spectrum of MSSM in left-right model visible sector brane construction
used in this paper.}
\label{SM Matter content}
\end{table}

The visible sector gauge groups can be broken down to the SM as
illustrated in \cite{Marchesano:2004xz}.
As pointed out in\cite{Marchesano:2004yq,Marchesano:2004xz}, the
visible sector is automatically supersymmetric
as long as the K\"ahler moduli parameters satisfy ${\cal A}_2={\cal A}_3$.
We can think of this as an entire plane in K\"ahler moduli space that
is supersymmetric for the visible sector.  It will actually be necessary
to impose this constraint to forbid giving vev's to
scalars charged under $SU(3)_{qcd}$, as we will see shortly.

The visible sector identified above cannot stand alone as it
does not cancel the RR tadpole conditions.  We need to introduce
a hidden sector for that purpose.
We will classify our choice of consistent hidden sector into two
categories: those that introduce one new NSNS tadpole constraint and those that introduce
two.
Of course, it is consistent to consider hidden sectors that introduce
more constraints.  Such constraints will generically not be satisfied, but
deformation of the brane embedding due to the veving of open string fields
will usually allow such solutions to preserve supersymmetry.  But the
addition of larger numbers of branes will make it more difficult for the
hidden sector to satisfy the RR-tadpole conditions.  For this reason (as well
as the computational difficulty in searching for solutions with many branes),
we will content ourselves in this work with finding hidden sectors that impose
at most two more constraints.  We will not attempt to study the non-perturbative
structure of the superpotential; the question of whether or not the
K\"ahler moduli are actually fixed~\cite{GarciadelMoral:2005js} 
in a supersymmetric solution thus remains open.

%%%%%%%%%%%%%%%%%%%%%%%%%%%%%%%%%%%%%%%%%%%%%%%%%%%%%%%%%%%%%%%%%%%%%%%%
\subsection{\label{one NSNS} Solutions with one NSNS tadpole constraint}

As noted above, the visible sector alone does not satisfy the
RR tadpole conditions.  The contributions from the branes of
the visible sector
to the $\vec Q$ vector are
\beq
\vec Q_{\rm vis}\equiv \sum_{k=a,b,c,d} N_k\vec Q(\Delta_k) = (72,8,2,2)
\eeq
We need the visible sector plus hidden sector of branes to cancel the RR
tadpole conditions
\beq
\vec Q_{\rm vis}+\vec Q_{\rm hid} + (32N_{\rm flux},0,0,0) = (16,16,16,16)
\eeq
which leads to the requirement that we find a hidden sector such that
\beq
\vec Q_{\rm hid}=(-56-32N_{\rm flux},8,14,14)
\eeq
In addition, we demand that the hidden sector preserve the same
$N=1$ supersymmetry as the visible sector.

The very large negative $Q^{\rm vis}_0$ charge needed is problematic.
Pure supersymmetric $D3$ branes or fluxes
only add positively to this charge and pure $D7$ branes
do not add to $Q_0$ at all.  What we need is a brane configuration with
large negative $Q_0$ and small or negative contributions to $Q_{1,2,3}$.
Furthermore, one can demonstrate from the NSNS tadpole conditions that a
brane can preserve supersymmetry only if at most one of its $Q$ charges is
negative\cite{Blumenhagen:2004xx}.
The only choice of brane that can preserve the same supersymmetry as the
orientifold and carry a negative charge is a magnetized $D9$ brane with
the appropriate signs in its wrapping numbers.

Note nevertheless that the addition of a hidden sector magnetized $D9$ brane
will introduce another NSNS tadpole condition
(applying eq.~\ref{supersymmetry condition} to the $D9$ brane).
Generically, any choice of $n$ magnetized
hidden sector branes will introduce $n$ additional NSNS tadpole constraints on the
K\"ahler moduli space.
Although it is not necessary to satisfy these constraints, it is useful
to classify solutions based on the number of constraints which arise from
the solution.  For every NSNS tadpole constraint which is not satisfied,
supersymmetry will require a scalar charged under the appropriate gauge group to
get a vev.  If such a scalar is also charged under the SM, then
this could be problematic for phenomenology.

The NSNS tadpole constraint of the visible would set ${\cal A}_2 ={\cal A}_3$,
which is a constraint\footnote{Note that this constraint appears
to only be a constraint on the real
K\"ahler moduli corresonding to the tori volumes, not on the axions which
complexify them.  How this constraint would be complexfied is an interesting
question~\cite{Douglas:1996sw}\cite{Cremades:2002te}.}
we assume the K\"ahler moduli satisfy in order to
minimize the risk of unacceptable gauge symmetry breaking (e.g.,
$SU(3)_{qcd}$ charged fields condensing).
When we add a hidden sector magnetized $D9$ brane to the theory
and demand that it be supersymmetric,
we are adding an additional FI-term constraint involving $\xi$, where
\beq
\xi \sim -\pi + \sum_i \tan^{-1}\left( \frac{|m_i|{\cal A}_i}{|n_i|}\right)
\eeq
subject to ${\cal A}_2={\cal A}_3$.  Therefore, we see that branes which are
related by $(n_2,m_2)\leftrightarrow (n_3,m_3)$ will impose the same
constraint on the K\"ahler moduli and thus can appear undeformed together in the
hidden sector while preserving supersymmetry.  A brane which is related
to another by such an interchange of winding numbers will be referred to
as a ``partner" brane.

We have constructed a computer program that searches for hidden sector
solutions that satisfy all the rules and constraints detailed in the previous
section and which are supersymmetric on 1D-surfaces in K\"ahler moduli space
(i.e., only one additional NSNS constraint in addition to ${\cal A}_2={\cal A}_3$).
After an exhaustive search we have identified six unique classes of
hidden sector solutions, listed in table~\ref{untilted models}.

\begin{table}[tbh]
\begin{center}
\begin{tabular}{cccccc}
\hline\hline %\vspace{0.1truein}
 & magnetized $D9$ brane & $N_{a}$ & $N_{\tilde a}$ & $N_{{D3},{D7_i}}$ &
$N^{\rm max}_{\rm flux}$ \\
\hline
1 & $(-2,1)(-3,1)(-4,1)$ & 2 & 2 & (40,0,0,0) & 1 \\
2 & $(-2,1)(-3,1)(-3,1)$ & 4 & - & (16,0,2,2) & 0 \\
3 & $(-2,1)(-2,1)(-7,2)$ & 2 & 0 & (0,0,0,6)  & 0 \\
4 & $(-2,1)(-2,1)(-7,2)$ & 0 & 2 & (0,0,6,0)  & 0 \\
5 & $(-2,1)(-2,1)(-5,1)$ & 2 & 2 & (24,0,0,0) & 0 \\
6 & $(-2,1)(-2,1)(-4,1)$ & 2 & 2 & (8,0,2,2)  & 0 \\
\hline\hline
\end{tabular}
\end{center}
\caption{The complete set of solutions to the hidden sector for the
SM embedding which are supersymmetric along a 1D-surfacee in K\"ahler
moduli space. Only the first model admits 3-form flux, and this model is equivalent up to
trivial sign reparametrizations to the one found by~\cite{Marchesano:2004xz}.}
\label{untilted models}
\end{table}

{}From these solutions we see that there is only one possible
solution with $N_{\rm flux}>0$ and that is obtained by adding one unit
of flux to the first solution in table~\ref{untilted models}.
We should note that our computer search for $N_{\rm flux}=1$ models
found 109 solutions when we do not take into account the K-theory constraint,
of which 65 allowed non-zero flux up to $N_{\rm flux}=10$.
Once we apply the K-theory constraint only this one solution given
above is left,
which through a trivial sign reparametrization is equivalent to the
model presented in~\cite{Marchesano:2004xz}.  There appear to be
no solutions satisfying all constraints
for $N_{\rm flux}\geq 2$ in these constructions that satisfy supersymmetry
on a 1D surface in K\"ahler moduli space.

These solutions have no net chiral exotics charged under $SU(4)_{PS}$.  This
feature is related to the automatic cancellation of the $SU(4)_{PS}$ cubic
anomaly, which in turn arises from the satisfaction of the RR tadpole
conditions\cite{Aldazabal:1999nu,Uranga:2000xp}.
The only contributions to the $SU(4)_{PS}$ cubic anomaly arise
from fermions transforming in the fundamental, anti-fundamental, symmetric
or anti-symmetric representations.
One can easily verify that no fermions transform in the symmetric or
anti-symmetric representation.  As a result, the number of fundamentals of
$SU(4)_{PS}$ must equal the number of anti-fundamentals, implying that
there are no net chiral exotics of this type.

%%%%%%%%%%%%%%%%%%%%%%%%%%%%%%%%%%%%%%%%%%%%%%%%%%%%%%%%%%%%%%%%%%%%%%
\subsection{Solutions with 2 NSNS tadpole constraints}

We may also consider solutions in which the hidden sector consists
of multiple branes which impose independent NSNS tadpole constraints.
If one adds no more than two NSNS tadpole constraints from the hidden
sector (in addition to the contribution from the visible sector), then
we might expect to be able to solve all constraints.  In fact, we will
find that although we have three constraints for three unknowns, the
constraints nevertheless cannot be solved simultaneously in most cases.

If we choose to add a hidden sector which generates 2 NSNS tadpole
conditions, then we must add two distinct hidden sector branes.  The
first must have negative D3-brane charge in order to cancel that RR
tadpole.  In order for it to be supersymmetric anywhere on moduli space,
it must therefore have positive values for all three D7-brane charges.

For the second brane, we have three choices.  First, we might choose
another brane with negative D3-brane charge and positive D7-brane charges.
Secondly, we might choose a brane with negative value for one D7-brane
charge, and positive values for all other charges.  Thirdly, we can
choose a brane with positive values for two of the four charges, and
with the other two charges being zero.  These are the only possibilities
which can be supersymmetric somewhere on K\"ahler moduli space, and which can
solve the RR tadpole conditions.

Again, we used a computer program to search for as many such solutions
as we could find. We performed a nearly exhaustive\footnote{We say
``nearly exhaustive'' because in this case, unlike the 1D K\"ahler
surfaces case of the previous subsection, we performed a randomized Monte
Carlo search for all solutions, and waited until it appeared all solutions
were obtained of the general structure we were searching.
However, we cannot guarantee all were found.} search through
all hidden sector brane configurations that would satisfy all constraints
and introduce 2 NSNS tadpole constraints.
Among the nearly 1000 classes of solutions we found,
we identified several solutions in which all NSNS tadpole
conditions can be solved simultaneously at a point (0D surface) in K\"ahler
moduli space.  In each of these cases, however,
no flux can be turned on, which means that we have not fixed the complex
structure moduli.  There is one case where all NSNS tadpoles can be solved
with $N_{flux}$=1, but in that case one of the K\"ahler moduli is infinite.
There were many more solutions we obtained where all such constraints
cannot be satisfied.

\begin{table}[tbh]
\begin{center}
\begin{tabular}{ccccccc}
\hline\hline %\vspace{0.1truein}
 brane $a$ & brane $b$ &  $N_{a}$ & $N_{b}$ & $N_{{D3},{D7_i}}$ &
$N^{\rm max}_{\rm flux}$ \\
\hline
 (-2,1)(-5,1)(-4,1) & (-2,1)(-2,1)(-1,1) & 2 & 2 & (0,0,0,4) & 1 \\
 (-4,1)(-4,1)(-4,1) & (0,1)(2,-1)(1,1) & 2 & 6 & (8,0,20,0) & 2 \\
 (-3,1)(-5,1)(-5,1) & (-1,1)(-1,1)(-1,1) & 2 & 2 & (0,0,2,2) & 3 \\
 (-4,1)(-6,1)(-4,1) & (0,1)(6,-1)(1,1) & 2 & 6 & (8,0,40,0) & 4 \\
 (-5,1)(-6,1)(-4,1) & (2,1)(1,-1)(2,-1) & 2 & 2 & (16,2,0,2) & 5 \\
 (-7,1)(-4,1)(-5,1) & (7,1)(2,-1)(1,-1) & 2 & 2 & (4,8,2,2) & 6 \\
 (-6,1)(-6,1)(-4,1) & (2,1)(1,-1)(2,-1) & 2 & 2 & (0,0,0,2) & 7 \\
 (-7,1)(-5,1)(-5,1) & (7,1)(1,-1)(1,-1) & 2 & 2 & (24,8,2,2) & 8 \\
 (-6,1)(-6,1)(-5,1) & (2,1)(1,-1)(2,-1) & 2 & 2 & (8,0,0,0) & 9 \\
(-7,1)(-5,1)(-5,1) & (3,1)(1,-1)(1,-1) & 2 & 2 & (0,0,2,2) & 9 \\
 (-6,1)(-5,1)(-6,1) & (4,1)(2,-1)(1,-1) & 2 & 2 & (0,4,0,0) & 9 \\
\hline\hline
\end{tabular}
\end{center}
\caption{Illustrative set of hidden sector solutions with $N_{flux}>0$ possible,
such that when combined with the visible sector of eq.~\ref{visible sector}
all RR tadpole constraints and K-theory constraints are satisfied. We have included
in this list all three $N_{flux}^{\rm max}=9$ solutions we have found.}
\label{two NSNS models}
\end{table}

Again, though, the solutions in which the NSNS tadpoles do not vanish are
not necessarily non-supersymmetric.  They instead are solutions in
which the simple brane configuration must be deformed in order for the
solution to become supersymmetric.  We have found many of these
models~\cite{web page},
of which we list in table~\ref{two NSNS models}
a representative for each value of $N_{flux}^{\rm max}$ from
1 to 9, which is the maximum value of $N_{flux}$ we obtained.  We list all
three $N_{flux}^{\rm max}=9$ solutions we found.

The existence of high flux $N_{flux}=9$ solutions is an
interesting result, because we know quite generally that the number of flux
vacua will grow rapidly with $N_{flux}$.  If the charge arising from fluxes is
much larger than the number of complex structure moduli, the number of
flux vacua will grow as ${N_{flux}^{2n+2}\over (2n+2)!}$.  If it is smaller, the number of
vacua is expected to grow
as $e^{\sqrt{2\pi (2n+2)N_{flux}}}$\cite{Ashok:2003gk}.
For all solutions we find, the exponential scaling is appropriate.  In
particular, for our highest flux model ($N_{flux}=9$) we find that the
number of vacua is approximately $10^{33}$.
Thus, a hidden sector with
the largest value of $N_{flux}$ will have the largest number of flux vacua, and
thus is most likely to have ``accidental" cancellations of phenomenological interest,
such as the cancellation of the bare cosmological constant~\cite{Bousso:2000xa} 
against its quantum correction.

It is important to note that we are not attempting to make a probabilistic prediction --
that a certain hidden sector is more likely to be chosen
by nature because it is realized in more flux vacua.  Instead, it is simply the
statement that a hidden sector with more flux vacua is more likely to exhibit one
flux vacua which, purely accidentally, happens to exhibit another property of
phenomenological interest, irrespective of whether or not nature chooses this
vacuum (for example, see \cite{statprob}).

%%%%%%%%%%%%%%%%%%%%%%%%%%%%%%%%%%%%%%%%%%%%%%%%%%%%%%%%%%%%%%%%%%%%%

\subsection{High Flux Model}

We would like to analyze some details of one of our $N_{flux}=9$ solutions
to give the reader a feel for the particle content and the symmetry breaking
patterns of this theory.  The brane content of the hidden sector
that goes along with the visible sector of eq.~\ref{visible sector} is
\bea
N_e=2 & (-6,1)(-5,1)(-6,1) \nonumber\\
N_f=2 & (4,1)(2,-1)(1,-1) \nonumber\\
N_g=4 & (1,0)(0,1)(0,-1)
\label{hidden sector}
\eea
The exotic states that have SM quantum numbers arise
from the intersections of hidden sector branes with
visible sector branes.  The resulting states will be
in the bifundamentals of the two gauge groups from
each respective brane.

In this case the chiral exotic matter is given by Table~\ref{exotic hiddens}.
For simplicity
we are overlapping the $U(3)$ and $U(1)_{B-L}$ brane stacks to form
a $U(4)=SU(4)\times U(1)_a$ stack.
We note that the total number of chiral exotics under $SU(3)$ is zero,
which is as it should be. Thus, all $SU(3)$ exotics have the chance
of obtaining mass, which is required for phenomenological viability.
The number of exotics under $SU(2)_L$ and
$SU(2)_R$ is even, which allows for the possibility of forming
$SU(2)_{L/R}$ gauge invariants to give mass to all of these exotics.

\begin{table}[tbh]
\begin{center}
\begin{tabular}{cccccc}
\hline\hline
Sector & $N_{\rm copies}$ & $SU(4)\times SU(2)_L\times SU(2)_R$ & $U(1)_a$ &
$U(1)_e$ & $U(1)_f$ \\
\hline
$(ae)$ & 24 &  ($\bar 4$,1,1) & -1 & 1 & 0  \\
$(ae')$ & 18 & (4,1,1) & 1 & 1 & 0 \\
$(af)$ & 10 &  (4,1,1) & 1 & 0 & -1 \\
$(af')$ & 4 & ($\bar 4$,1,1) & -1 & 0 & -1 \\
$(be)$ & 36 & (1,2,1) & 0 & 1 & 0 \\
$(bf)$ & 4 & (1,2,1) & 0 & 0 & -1 \\
$(ce)$ & 30 & (1,1,2) & 0 & 1 & 0 \\
$(cf)$ & 8 &  (1,1,2) & 0 & 0 & -1\\
\hline
$(ef)$ & 150 & (1,1,1) & 0 & -1 & 1 \\
$(ef')$ & 98 & (1,1,1) & 0 & 1 & 1 \\
$(eg)$ & 30 & (1,1,1) & 0 & 1 & 0 \\
$(fg)$ & 2 & (1,1,1) & 0 & 0 & 1 \\
$(ee')$ & 530 & (1,1,1) & 0 & $- 2$ & 0 \\
$(ff')$ & 54 & (1,1,1) & 0 & 0 & $- 2$ \\
\hline\hline
\end{tabular}
\end{center}
\caption{We list in this table all exotic chiral multiplets charged
jointly under a SM gauge group and a hidden sector gauge group for the
$N_{flux}=9$ model discussed in the text.}
\label{exotic hiddens}
\end{table}

In particular, to see how all the exotics can get mass,
we might give vevs to the scalar of one of the
98 chiral multiplets charged under $U(1)_e$ and $U(1)_f$, with charge
1 under both groups.  This amounts to the brane recombination
\beq
[e] + [f'] \rightarrow [j]
\eeq
and will leave us with 56 $SU(4)$ exotics, 40 $SU(2)_L$ exotics and
38 $SU(2)_R$ exotics.  There remain scalars transforming in the
symmetric representation of the gauge theory living on $[j]$, and
giving a vev to these scalars corresponds to the further brane recombination
\beq
[j] + [j'] \rightarrow [k]
\eeq
This removes all chiral $SU(3)$ and
$SU(2)_{L,R}$ exotics.

Alternatively, we could have begun by giving vevs to scalars transforming
under $U(1)_e$ and $U(1)_f$ with charges 1 and -1 respectively.  This
corresponds to the brane recombination
\beq
[e] + [f] \rightarrow [j]
\eeq
and leaves us with 28 $SU(4)$ exotics, 32 $SU(2)_L$ exotics and
22 $SU(2)_R$ exotics.  Again, the recombination $[j]+[j']
\rightarrow [k]$ removes all chiral exotics.
It is interesting to note that to eliminate the chiral
exotics, it is necessary to introduce two scales of symmetry breaking
in the hidden sector in addition to electroweak symmetry breaking in the visible sector.
This is a generic consequence of any hidden sector with two
additional NSNS tadpole constraints.  More hidden sector branes
would in general add more $D$-term constraints whose solution may
require more symmetry breaking scales. It would be interesting
to investigate the model-building and observable
implications of multiple symmetry breaking scales.

What's most interesting about this model is the high amount of flux.
As stated above, the number of flux vacua is $\sim 10^{33}$, multiplied
by other prefactors.
One such prefactor arises from the
integration of the vacuum density over the complex structure moduli
space.  Another arises from the fact that only a fraction of the vacua
found here will have moduli that are stabilized in a self-consistent
regime (i.e., small coupling and volume larger than string scale).
If we assume that these prefactors are not too small, we can still
easily obtain at least one vacuum state with cosmological constant
nonzero but at or below the current measured value;
we need of ${\cal O}(10^{30})$ vacua for that.  If the number of vacua
is significantly larger than ${\cal O}(10^{30})$, it is possible that one
could also have accidental fine-tuning of other observables to be consistent
with current experiment or
theoretical prejudice, such as gauge coupling unification,
cold dark matter abundance, acceptable CP violating
phases, etc.\
This is what makes high $N_{flux}$
vacua that we have found especially interesting, since landscape statistics
has a chance of enabling some good features of the model to be present
simultaneously in at least one vacuum.

%%%%%%%%%%%%%%%%%%%%%%%%%%%%%%%%%%%%%%%%%%%%%%%%%%%%%%%%%%%%%%%%%%%%%
\section{\label{phenomenological considerations} Phenomenological Considerations}

Our goal in the above was to construct SM embeddings within type IIB
flux compactifications.  We were drawn to the framework
of $T^6/Z_2\times Z_2$ orientifold compactification
with $SU(3)\times SU(2)_L\times SU(2)_R\times U(1)$
gauge group, partly due to its simplicity.  The overall framework
puts restrictions on the phenomenology, some of which are challenges
to phenomenological viability.

Likewise, if we were to focus primarily on low-energy effective theory
model building, with little consideration for how such a model could be
embedded in a more complete framework like string theory, we would likely
arrive at models that are challenges to string model viability.  (The
pure SM is of course one such model.)
In this section we discussion some of the phenomenological implications
of the framework we have detailed above, with the goal of gaining
further understanding of what aspects are viable from both the
phenomenological and string perspectives.

\noindent
{\em Exotic Gauge Symmetries}

Additional gauge symmetries beyond the SM are common in many approaches
to string model building. There is no exception here, and
the general reason for this in our case is that
right-handed leptons
in the SM spectrum are charged only under hypercharge $U(1)_Y$
and not $SU(3)$ color or weak $SU(2)_L$.  Since matter states arise
from their intersections between brane stacks, yielding bifundamental
representations, we need a fourth brane stack to have intersection
with a ``hypercharge stack'' (or equivalent) to get right-handed
leptons.  Therefore, there is always need for some exotic gauge
symmetry arising from this fourth stack.  Additional gauge symmetries
can arise from more hidden sector brane stacks

There is a substantial body of literature~\cite{Leike:1998wr} on the phenomenology
of gauge bosons associated with exotic non-SM symmetries at both
the TeV scale and the intermediate scale.  At the
level of our model building, we take no position on what scale extra
symmetries are most likely to appear.  We only note that their
presence is required somewhere in the energy continuum.

In addition to normal extra $U(1)$ gauge symmetries,
the generalized Green-Schwarz
mechanism in these theories enables the possibility of having a low-scale
global symmetry exactly preserved, with perhaps some small breaking due
to a small vev, whereas the corresponding gauge bosons of the symmetry
are very massive.  The large mass comes from a Stueckelberg term in the
potential, that can be seen after the shift is made to cancel the
symmetry's anomaly. A $U(1)$ symmetry with a Stueckelberg mass can
be considered somewhat generic in this framework, and the phenomenology
of this case is unique and rich~\cite{Kors:2005uz}.

\noindent
{\em Supersymmetric unification}

The theories we have analyzed above are supersymmetric, in the sense that
each brane stack satisfies the NSNS tadpole constraint and preserves
supersymmetry somewhere in K\"ahler moduli space.  Supersymmetry breaking
can occur through a myriad of possibilities, but each model has the
prospect of supersymmetry breaking~\cite{susy breaking}
giving rise to soft masses anywhere
between the weak scale and the Planck scale.  Thus, softly broken
supersymmetry is a phenomenological implication of this scenario,
but it is not yet clear at what scale it should be found.

One of the attractive features of low-energy supersymmetric theories
is the apparent unification of gauge couplings
if the three gauge couplings of the minimal supersymmetric SM
are renormalization group evolved up to the high scale. This may be a profound
clue to nature or an interesting accident.  In any event,
being a theory with different brane stacks for different gauge groups,
gauge coupling unification is by no means automatic or expected
in our type IIB theories discussed above.

Pati-Salam unification is possible if we allow both the $U(3)$ and $U(1)$
brane stacks to be on top of each other.  However, the scale at which
the $SU(4)_{\rm PS}$ is recovered is a model building, or rather a
model analyzing, detail that is as yet unknown.  Nevertheless, it would
be of interest to consider partial unification of the SM
into $SU(4)$ as a generic phenomenological outcome, and see what restrictions
this would have on the spectrum\footnote{Tests of all unification propositions must
input the known masses and couplings, which then often puts strong
restrictions on the unknown masses and couplings of the model.}
from a bottom up point of view.

\noindent
{\em Gauge symmetry breaking, $R$-parity and neutrino masses}

One appealing way to break $SU(2)_L\times SU(2)_R\times U(1)_{B-L}$ down
to the SM is to first condense a $(1,3,2)$ field to break
$SU(2)_R\times U(1)_{B-L}\to U(1)_Y$ and then condense a $(2,2,0)$ bidoublet
field to break $SU(2)\times U(1)_Y$ down to $U(1)_{em}$.  The advantages
of this approach are that the seesaw mechanism can induce small neutrino
masses and $R$-parity can be retained as a discrete $Z_2$ subgroup of
$U(1)_{B-L}$~\cite{Mohapatra:1998au}.

The spectrum allowed in the visible sector of our $D$-brane configuration
does not allow this symmetry breaking pattern.  Instead, the first step
of the symmetry breaking pattern is accomplished through an $SU(2)_R$
doublet field, such as the right-handed slepton.  Indeed, an equivalent
description is that of brane recombination
of the $c$ and $d$ branes, which is equivalent to
veving a field charged under the two branes
along a flat direction~\cite{Marchesano:2004xz}.

It is possible to generate viable neutrino mass spectra even restricting
ourselves to the matter content available to us, by using one of the
many approaches within left-right model building~\cite{Ma:2003zy}.
However, for symmetry breaking accomplished by right-slepton
field condensation, $R$-parity is spontaneously broken since
an odd lepton-number is being carried into the vacuum.  Although
this does not allow the lightest supersymmetric particle to be
the dark matter in a straightforward way, dark matter could arise from another source
such as vector-like matter with their own discrete symmetry
or axions. Proton decay is also not automatically a problem,
as only lepton number is spontaneously broken.  Baryon number
can stay preserved from the perspective of this gauge symmetry
breaking pattern.

%%%%%%%
\noindent
{\em Supersymmetry breaking}

In the previous sections we have searched for supersymmetric vacua
which admit SM gauge group and matter content, but have
also discussed some aspects of supersymmetry breaking.  In any
realistic string vacuum, experiment tells us that we have no moduli
and no supersymmetry.  In general, we would expect that supersymmetry
breaking effects will generate potentials for generic scalars.  But
these potentials can very easily destabilize our solution, and in
general it is quite difficult to maintain enough control over the
calculation to be sure that we really have nonsupersymmetric
solutions (e.g., see~\cite{Kachru:2002gs}).
Thus, one of the central ideas of flux vacua counting has
been to find supersymmetric solutions with no moduli, and then add
supersymmetry breaking effects at a lower scale.  If moduli can be
fixed in a supersymmetric compactification, then we certainly know
that our solution is stable.  When we then add supersymmetry breaking
effects at a lower scale, we can be confident that the solutions is
not destablized because the various scalars have already been given a
mass at a much higher scale.  The solution may be deformed slightly by
the supersymmetry-breaking effects, but it could not be destabilized.

A fundamental point in our search for SM flux vacua is
that this entire story cannot proceed as before: if we want to get the
SM, we cannot fix all moduli before breaking supersymmetry.  The
matter content of the SM will include chiral fermions.  Their
superpartners will be moduli unless supersymmetry is broken.  As a
result, any type of counting which focusses on vacua with no moduli
must input supersymmetry breaking.

Although we cannot fix all moduli at a very high scale, we can fix many closed
string moduli.  This is good, because in general it has proved harder to understand
whether potentials for these moduli destabilize solutions or simply deform them.  On
the other hand, instabilities involving open string moduli tend to resolve themselves
by the annihilation of branes, without the destabilization  of the space-time
compactification itself.

But it will be true that the stabilization of at least some K\"ahler moduli
will be inextricably linked to supersymmetry breaking.  We
require ${\cal A}_2={\cal A}_3$ in order to
avoid undesirable deformations of the SM branes, but
we generically cannot expect that this condition will arise from
the $F$-term equations.  Instead, we must demand that soft-masses arising
from supersymmetry breaking give masses to the squarks and sleptons, with
the equations of motion arising from the $D$-term potential(\ref{Dtermpotential})
then fixing ${\cal A}_2={\cal A}_3$.
This suggests that at least some of the K\"ahler moduli might receive a mass
which, like the squark and slepton masses, is set by the supersymmetry breaking
scale.

Due to these uncertainties, we have not focussed on any particular method of
supersymmetry breaking.  On the other hand, since it
is an important part of the overall puzzle, we will list
mechanisms by which supersymmetry
breaking can be achieved, how the different approaches
can be integrated into the string framework,
and what kind of phenomenological implications they have.

The most obvious way to break supersymmetry is to add a badly misaligned
brane that cannot be supersymmetric in the K\"ahler moduli space.  The quintessential
example of this is a $\overline{D3}$ brane.  Supersymmetry breaking is at
tree-level and is generically of order the string scale in this case.
Since they contribute negatively to the $D3$-brane charge, one would be tempted
to assume that it is possible to
add an arbitrary number of $\overline{D3}$ branes with an arbitrary amount
of flux. However, this is not the case, as the configuration is unstable to
$\overline{D3}$ annihilations with flux~\cite{Kachru:2002gs}. Furthermore,
given the considerations
of ref.~\cite{Kachru:2002gs}, which translates to $N_{\overline{D3}}\lsim 3 N_{flux}$,
it is not possible to start with an $N_{flux}=0$ model of sec.~\ref{one NSNS}
and construct a $N_{flux}=1$ model with added $\overline{D3}$ branes.

Nevertheless, it is possible to add a $D3/\overline{D3}$ combination to
any $N_{flux}>0$ model and obtain supersymmetry breaking.  It is expected
that the $\overline{D3}$ will want to locate in a warped throat, thereby
possibly suppressing what would otherwise by a high-scale supersymmetry
breaking mass terms to the weak scale.  The supersymmetry breaking in
this case would be $D$-term and could lead to hierarchically larger
scalar masses than gaugino masses.

IASD and ISD(0,3) fluxes\footnote{For ISD(0,3) fluxes, the generation of masses is
less certain, as it depends on how the no-scale structure is
broken (see Camara et al.\ in~\cite{susy breaking}).  Of course,
IASD fluxes will also induce F-terms for complex structure moduli.}
could both contribute to $F$-term-like supersymmetry breaking,
which gives rise to soft masses of open
string states connecting $D7$ branes to other branes.  Both IASD and ISD
supersymmetry breaking
can in principle be small, and lead to weak scale supersymmetry breaking even
with the standard large hierarchy of the string scale and weak scale.

We remark that in the case of ISD/IASD fluxes contributing to supersymmetry breaking,
the FI $D$-terms arising from misaligned supersymmetric branes in K\"ahler moduli
space will no longer be able to zero themselves out completely by
appropriate choice of scalar vevs, and so they will contribute to the overall supersymmetry
breaking accounting in the low-energy phenomenology.
One should think of these $D$-terms as simply constraints that involve both
the K\"ahler moduli and open-string moduli.  Given other constraints on the K\"ahler
and open-string moduli arising from $F$-terms, it may not be possible to simultaneously
solve the $F$-term and $D$-term equations.
However, the size of that supersymmetry
breaking is controlled by the $F$-terms: as we tune the $F$-terms to zero the FI-induced
$D$-terms must go to zero.

Lastly, gaugino condensation is another potential source of supersymmetry breaking
whenever the hidden sector has a large enough gauge group and sufficiently small
matter content.  This supersymmetry breaking is $F$-term.  Although gaugino condensation
is not necessarily a crucial ingredient of supersymmetry breaking (fluxes can
do all the work), it might nevertheless play an important role in fixing K\"ahler moduli
through its non-perturbative dynamics.

%%%%%%%%%%%%%%%%%%%%%%%%%%%%%%%%%%%%%%%%%%%%%%%%%%%%%%%%%%%%%%%%%%%%%%%%
\section{Conclusions}

One of the major lessons of this exercise has been the importance of the
exact method by which the open string and K\"ahler moduli are fixed.  We have
seen that this is linked to the mechanism by which supersymmetry is broken.
In SM flux vacua of the type we discuss, the masses given to complex
structure moduli and the dilaton can be made numerically much
larger than the
scale of supersymmetry breaking.  But the masses given to K\"ahler moduli may
be of the same scale as those given to open-string moduli, and these
are of the order of the supersymmetry breaking scale.  Thus it seems that these
three problems (breaking supersymmetry, fixing K\"ahler moduli and fixing
open string moduli) may have to be dealt with simultaneously.

One can consider these various types of hidden sectors, and under the assumption
that any embedding leaves all K\"ahler moduli fixed (without overconstraining
the moduli),
compare the number of flux vacua.  In that case, as follows our intuition from
flux vacua counting\cite{Ashok:2003gk,Denef:2004ze},
we see that the number of vacua scales as
$e^{\sqrt{2\pi (2n+2)N_{flux}}}$
where $n=51$ is the number of complex structure moduli and $N_{flux}$ is the
amount of three-form flux turned on as part of the hidden sector.
We thus see that we
can get insight into the nature of flux vacua counting in this theory, but to
have complete control over the counting, we should understand the
nature of the
non-perturbative corrections to the superpotential.
Furthermore, it would be interesting to
compare the number of flux vacua with this SM embedding to the total
number of flux vacua for this orientifold compactification.  Perhaps one
can estimate
the total number of flux vacua using techniques similar to those used in
\cite{Blumenhagen:2004xx}.

Our discussion has thus far been for a choice of discrete torsion which
yields 51 complex structure moduli and 3 K\"ahler moduli.  However,
for a different choice of discrete torsion ($B$-field turned
on the shrunken cycles)\cite{Gopakumar:1996mu}, we would instead find
3 complex structure moduli and
51 K\"ahler moduli.  This would significantly reduce the scaling of the
number of flux vacua with $N_{flux}$.  In this case, we would have
many K\"ahler moduli that can participate in constraints arising
from non-perturbative
superpotential corrections, but not from NSNS tadpole constraints (which only
affect the 3 ``toroidal" K\"ahler moduli).  This may affect how easy it is to
fix all the K\"ahler moduli, even if one has the right number of constraints
(i.e., the toroidal set of 3 might be over-constrained, while the
non-toroidal
set of 48 is under-constrained).

We have identified many solutions with one or two NSNS tadpole constraints.
It would be interesting to do a systematic search
for hidden sectors with three or more NSNS tadpole constraints.  As we discussed
in the text, such solutions are likely to be much more difficult to find,
as higher numbers of branes will struggle to satisfy the RR tadpole constraints.
Nevertheless, one should search for their existence, and if there are solutions,
determine their $N_{flux}^{\rm max}$.

Another subtlety which we do not address is the possibility of wrapping
branes on the shrunken cycles of the orientifold.  Although we do not know if
such branes can participate in a SM embedding, they may affect
flux vacua counting within our choice of orientifold.

The approach we have used in this study 
can be generalized to other string compactifications beyond
$T^6/Z_2 \times Z_2$. The important data needed for analysis of the theory
are the intersection numbers and the D-brane charges, and how they interact
with the tadpole constraints.  In our case, the analysis was simplified by
manipulating the brane wrapping numbers.  In other compactifications
different manipulation techniques are required, but the general
procedure is the same.

%%%%%%%%%%%%%%%%%%%%%%%%%%%
\section*{Acknowledgements}
We gratefully acknowledge R. Blumenhagen, M. Douglas, S. Kachru,
S. Sethi and G. Shiu for useful discussions. This work has been
supported in part by the Department of Energy and the Michigan
Center for Theoretical Physics (MCTP). We also thank the
Aspen Center for Physics.

\end{document}